\documentclass[]{aastex631}

\begin{document}

\title{Infrared Cavity Surrounding T Pyxidis: Evidence of a Nova Super-remnant?}

\author[0000-0001-6337-6871]{Michael W. Healy-Kalesh}
\affiliation{Astrophysics Research Institute, Liverpool John Moores University, IC2 Liverpool Science Park, Liverpool, L3 5RF, UK}
\author[0000-0003-0156-3377]{Matthew J. Darnley}
\affiliation{Astrophysics Research Institute, Liverpool John Moores University, IC2 Liverpool Science Park, Liverpool, L3 5RF, UK}

\begin{abstract}
\noindent Nova super-remnants (NSR) are shell-like structures spanning tens of parsecs that are predicted to grow around all recurrent novae as a result of repeated ejecta sweeping up the surrounding material. After the first NSR was discovered in the Andromeda Galaxy around M\,31N 2008-12a, a number of Galactic NSRs have been uncovered, as has the first NSR in the Large Magellanic Cloud. Here we report an apparent infrared cavity surrounding the Galactic recurrent nova, T Pyxidis, which we suggest may be a nova super-remnant and discuss the implications.
\end{abstract}

\section{Introduction} \label{sec:Introduction}
\noindent Nova super-remnants (NSRs) are substantial shells made up almost exclusively of interstellar medium (ISM) swept up by recurrent nova (RN) eruptions over millenia \citep{2024PoS...460...35H}. The prediction that NSRs should exist around all RNe \citep{2023MNRAS.521.3004H} is now being validated: the prototype NSR associated with the Andromeda Galaxy RN M31N 2008-12a \citep{2019Natur.565..460D} has been joined by NSRs hosted by the Galactic RNe KT Eridani \citep{2024MNRAS.529..224S}, T Coronae Borealis \citep{2024ApJ...977L..48S}, RS Ophiuchi \citep{2025AJ....170...56S} and V745 Scorpii (Healy-Kalesh et al. submitted), as well as the Large Magellanic Cloud RN, LMCN 1971-08a \citep{2025A&A...702L...9H}.

\section{Nova super-remnant around T Pyxidis} \label{sec:Nova super-remnant around T Pyxidis}
\noindent T Pyxidis is one of the eleven known Galactic RNe \citep{2021PoS...368...44D,2024MNRAS.529..224S}, with eruptions recorded in 1890, 1902, 1920, 1944, 1967, and 2011 \citep{2021PoS...368...44D}. Such relatively quick-fire eruptions result from a combination of its high mass white dwarf \citep[${\sim}1.33$\,M$_{\odot}$;][]{2026ApJ..1000..219S} and high accretion rate \citep[${\sim}10^{-7}$\,M$_{\odot}$yr$^{-1}$;][]{2026ApJ..1000..219S}. Though, T\,Pyx is unusual in comparison to many other RNe on account of its short orbital period \citep[1.83 hr;][]{1998PASP..110..380P} and is unique as being the only RN hosting a persistent nova shell \citep{2015ApJ...805..148S}.

\citet{2026ApJ..1000..219S} proposed that the current RN cycle of T Pyx was triggered with a classical nova eruption in ${\sim}1866$ and likely ended with the 2011 eruption, and that the nova will now enter a ${\sim}13,000$\,yr phase of quiescence. Orbital period variations across multiple eruptions may indicate that the T Pyx white dwarf (WD) is decreasing in mass \citep{2026ApJ..1000..219S}. Yet modelling of NSRs \citep{2023MNRAS.521.3004H} indicates that neither of these properties of T\,Pyx would prevent the creation of an NSR.

\section{Infrared data of T Pyxidis surroundings} \label{Infrared data of T Pyxidis surroundings}
\noindent Based on a suite of NSR simulations provided in \citet{2023MNRAS.521.3004H}, it is predicted that (i) novae with high accretion rates similar to T Pyx (${\sim}10^{-7}$\,M$_{\odot}$yr$^{-1}$) are favourable for growing NSRs, with sizes ranging $20-200$\,pc; (ii) systems with eroding WDs can also grow NSRs (this effect was illustrated with a lower accretion rate); and (iii) systems with recurrence periods of ${\sim}3000$\,yr and ${\sim}49,000$\,yr are still capable of growing vast NSRs. Therefore, we predict that an NSR {\it could} be created by T Pyx and, if so, would have a size of the same order of magnitude as the NSRs around other Galactic RNe (${\sim}30 - 200$\,pc). In addition, previous searches for evidence of an NSR being present around other Galactic RNe (\citealt {2024MNRAS.529L.175H}, Healy-Kalesh et al. submitted) were conducted using the new generation {\it Infrared Astronomical Satellite} ({\it IRAS}) catalogue, known as IRIS \citep{2005ApJS..157..302M}. 

\begin{figure}[htbp]
    \centering
    \includegraphics[width=\linewidth]{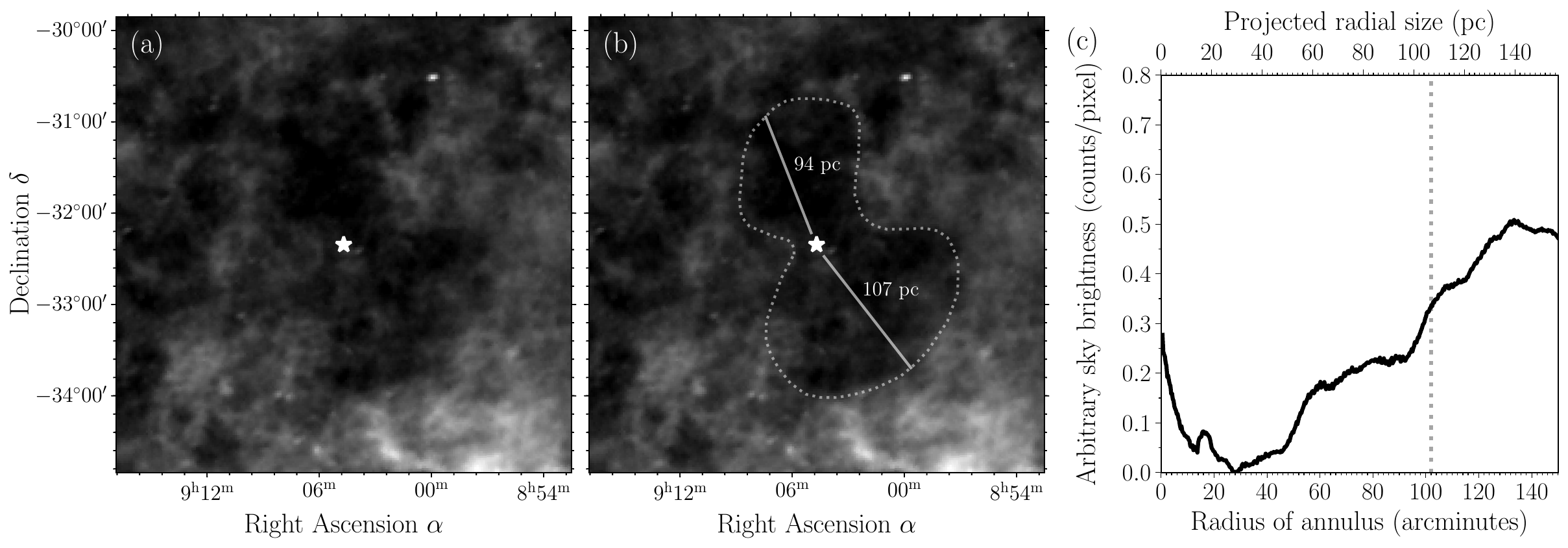}
    \caption{Panel (a) $-$ $5^{\circ}\times5^{\circ}$ IRIS 60\,$\mu$m far-infrared image of the surroundings of T Pyxidis, with the nova's location indicated with a white star. Panel (b) $-$ Same as panel (a) but with a boundary of the apparent cavity outlined with a dashed line for guidance. Panel (c) $-$ Arbitrary surface brightness radial profile of the surroundings of T Pyxidis from annular photometry (see text). The largest extent of the apparent cavity at 107 pc is indicated with the dashed line.}
    \label{NSR cavity}
\end{figure}

With the predicted approximate size and the success of previously identifying possible NSRs in IRIS data, we inspected the extended surroundings ($5^{\circ} \times 5^{\circ}$ FOV) of T Pyx within IRIS 60\,$\mu$m imaging and identified a large cavity coincidentally aligned with the nova, with T Pyx centrally located, as shown in the left panel of Figure~\ref{NSR cavity}.

\section{Discussion\label{sec:Discussion}}
\noindent The cavity appears to be bilobed (middle panel of Figure~\ref{NSR cavity}), with the larger southwest component spanning ${\sim}102^{\prime}$ and the smaller northeast component spanning ${\sim}90^{\prime}$. At the distance of ${\sim}3.6$\,kpc to T\,Pyx \citep{2026ApJ..1000..219S}, the southwest and northeast component would have scales of ${\sim}107$\,pc and ${\sim}94$\,pc, respectively, thereby conforming to the scale of known NSRs. Such an NSR with a bilobed structure is reminiscent of the NSRs (found through their H$\alpha$ emission) around T CrB \citep{2024ApJ...977L..48S} and RS Oph \citep{2025AJ....170...56S}.

In addition to the visual evidence of a cavity around T Pyx in the IRIS data, we also quantify this lack of emission. Using the method presented in \citet{2024MNRAS.528.3531H}, we performed annular photometry on the IRIS data by employing 500 circular annuli centred on the location of T\,Pyx with radii ranging from $0.375-150$ arcminutes, with the results provided in the right panel of Figure~\ref{NSR cavity}. As expected, this preliminary analysis reveals a general profile consisting of low levels of emission (cavity) bordered by higher levels of emission (surrounding medium). Further, the profile can be separated into varying components attributable to the structure seen in the data: moderately high levels of emission close to the nova ($ \lesssim8$\,pc) before a drop to negligible levels between ${\sim}20-45$\,pc. The emission then increases at around 60\,pc (likely due to the influence of the `pinched' part of the cavity) before another substantial increase at around 100\,pc, coinciding with the approximate edges of the two lobes.

We also note that the same cavity feature has been previously independently identified in \citet{2008ASPC..401...90V}, though a much smaller diameter of $2^{\prime}$ is specified. In this work, it was proposed that this cavity may have formed more than $10^4$ years ago and ``possibly built up during successive eruptions'' \citep{2008ASPC..401...90V}. In light of the first NSR being discovered in M\,31 in 2015 \citep{2015A&A...580A..45D,2019Natur.565..460D} and subsequent development of both the theoretical and observational study of this phenomena, we corroborate the suggestion that the cavity seen around T\,Pyx was likely created through the excavation of local ISM by repeated nova eruptions in the form of an NSR, though we would suggest a longer evolutionary history based on studies of other NSRs \citep{2024MNRAS.529..236H,2024ApJ...977L..48S,2025AJ....170...56S}.  

With the existence of an infrared cavity revealing the location and extent of the possible NSR surrounding T\,Pyx, it would be fruitful to search the extended environs (up to ${\sim}100^{\prime}$) of the RN in other wavebands, particularly with a narrowband H$\alpha$ filter as NSRs have been shown to emit strongest in H$\alpha$ \citep{2019Natur.565..460D,2024MNRAS.529..224S,2024ApJ...977L..48S,2025AJ....170...56S,2025A&A...702L...9H}, to confirm the presence of an NSR around T Pyx. Furthermore, the implications of an existing NSR possibly cast doubt on the long-term past and future behaviour of T Pyx and suggest that (perhaps) it behaves more like the `typical' RNe that has previously been proposed.

\vspace{5mm}
\facilities{{\it Infrared Astronomical Satellite} ({\it IRAS})}

\newpage
\bibliography{bibliography}{}
\bibliographystyle{aasjournal}

\end{document}